\documentclass[a4paper,11pt]{article}
\usepackage{epsf}
\usepackage{gdgspace}
\usepackage{amssymb}
\usepackage{chicago}
\usepackage{geometry}
\newcommand{\be}{\begin{equation}}
\newcommand{\ee}{\end{equation}}
\newcommand{\bd}{\begin{displaymath}}
\newcommand{\ed}{\end{displaymath}}
\newcommand{\bea}{\begin{eqnarray}}
\newcommand{\eea}{\end{eqnarray}}
\newcommand{\bi}{\begin{description}}
\newcommand{\ei}{\end{description}}
\newcommand{\bq}{\begin{quote}}
\newcommand{\eq}{\end{quote}}

\def\ni{\noindent}

\def\a{\alpha}
\def\b{\beta}

\def\C{\Chi}
\def\g{\gamma}
\def\G{\Gamma}
\def\d{\delta}
\def\D{\Delta}

\def\t{\tau}

\def\m{\mu}
\def\n{\nu}

\def\l{\lambda}
\def\L{\Lambda}
\def\s{\sigma}
\def\p{\phi}

\def\Z{{\sf Z\kern-4.5pt Z}}
\def\R{{\sf R\kern-5.5pt I}}
\def\Q{{\sf C\kern-5.0pt Q}}
\def\C{{\sf C\kern-5.0pt C}}
\def\v{\vspace{1.0cm}}
\begin{document}
\author{Alexander~Unzicker and Timothy~Case\\
        Pestalozzi-Gymnasium  M\"unchen\\[0.6ex]}
\title{Translation of Einstein's Attempt of a\\ 
Unified Field Theory with Teleparallelism}
\date{March 6th, 2005}
\maketitle

\begin{abstract}
We present the first English translation of Einstein's original 
papers related to the teleparallel\footnote{`absolute parallelism', 
`distant parallelism' and the German `Fernparallelismus' are synonyms.}
attempt of an unified field theory of gravitation and electromagnetism.
Our collection contains the summarizing paper in 
 Math. Annal. 102 (1930) pp.~685-697 and 2 reports published
in 'Sitzungsberichte der Preussischen Akademie der Wissenschaften'
on June 7th, 1928 (pp.~217-221), June 14th, 1928 (pp.~224-227) 
and a precursor report (July 9th, 1925 pp.~414-419).
To ease understanding, literature on tensor analysis is
quoted in the footnotes. 
\end{abstract}

\setcounter{secnumdepth}{0}

\newpage
\setcounter{section}{0}
\setcounter{equation}{0}
\setcounter{page}{1}

\v
\begin{center}
\huge{Unified Field Theory of Gravitation and Electricity\\}
\v
\Large
Albert~Einstein \\

\normalsize
translation by A.~Unzicker and T.~Case\\
\v
\large{Session Report of the Prussian Academy of Sciences, pp.~414-419\\
July 25th, 1925}
\end{center}
\v

Among the theoretical physicists working in the field of the
general theory of relativity there should be a consensus about the
consubstantiality of the gravitational and electromagnetic field.
However, I was not able to  succeed in finding a convincing
formulation of this connection so far. Even in my article
published in  these session reports (XVII, p.~137, 1923) which is
entirely based on the foundations of {\sc Eddington}, I was of the
opinion that it does not reflect the true solution of this
problem. After searching ceaselessly in the past two years I think
I have now found the true solution. I am going to communicate it
in the following.

The applied method can be characterized as follows.
First, I looked for the formally most simple expression for the law of gravitation in
the absence of an electromagnetic field, and then the most natural generalization
of this law. This theory appeared to contain  {\sc Maxwell}'s theory in first approximation. In the
following I shall outline the scheme 
of the general theory ($\S~1$) and then show in which sense
this contains the law of the pure gravitational field ($\S~2$) and {\sc Maxwell}'s theory ($\S~3$).

\section{$\S~1.$ The general theory}
 We consider a 4-dimensional continuum with an
affine connection, i.e. a $\G_{\a \b}^{\m}$-field which defines
infinitesimal vector shifts according to the relation \be d
A^{\m}= - \G_{\a \b}^{\m} A^{\a} dx^{\b}. \ee We do not assume
symmetry of the $\G_{\a \b}^{\m}$ with respect  to the indices
$\a$ and $\b$. From these quantities $\G$ we can derive  the
{\sc Riemann}ian tensors

\bd
R_{\m  . \n \b}^{\a} =
- \frac{\partial \G_{\m \n }^{\a}}{\partial x_{\b}}+
\G_{\s \n}^{\a} \G_{\m \b}^{\s}
+ \frac{\partial \G_{\m \b }^{\a}}{\partial x_{\n}}+
 \G_{\m \n}^{\s} \G_{\s \b}^{\a}
\ed
 and
\be
R_{\m \n}=
R^{\a}_{\m  . \n \a} = - \frac{\partial \G^{\a}_{\m \n }}{\partial x_{\a}}+
\G_{\m \b}^{\a} \G_{\a \n}^{\b}
+ \frac{\partial \G^{\a}_{\m \a }}{\partial x_{\n}}+
\G_{\m \n}^{\a} \G_{\a \b}^{\b}
\ee
in a well-known manner.
Independently from this affine connection we introduce a contravariant
tensor density $\mathfrak{g}^{\m \n}$, whose symmetry properties we leave undetermined as well. 
From both quantities we obtain the scalar density
\be
\mathfrak{H} = \mathfrak{g}^{\m \n} R_{\m \n}
\ee
and postulate that all the variations of the integral
\bd
{\cal J} = \int \mathfrak{H} \ d x_1 d x_2 d x_3 d x_4
\ed
with respect to the $\mathfrak{g}^{\m \n}$ and $\G_{\m \n}^{\a}$ as independent (i.e.
not to be varied at the boundaries) variables vanish.

The variation with respect to the $\mathfrak{g}^{\m \n}$ yields the 16 equations
\be
R_{\m \n} =0,
\ee
the variation with respect to the $\G_{\m \n}^{\a}$ at first the 64 equations
\be
\frac{\partial \mathfrak{g}^{\m \n }}{\partial x_{\a}}+ \mathfrak{g}^{\b \n} \G_{\b \a}^{\m}+
\mathfrak{g}^{\m \b} \G_{\a \b}^{\n}-
\d_{\a}^{\n} \Bigl( \frac{\partial \mathfrak{g}^{\m \b }}{\partial x_{\b}}+
 \mathfrak{g}^{\s \b} \G_{\s \b}^{\m} \Bigr)
-\mathfrak{g}^{\m \n} \G_{\a \b}^{\b}=0.
\ee


We are going to begin with some considerations that allow us to replace the eqns.~(5)
by simpler ones. If we contract the  l.h.s. of (5) by $\n$ and $\a$ or
$\m$ and $\a$, we obtain the equations

\be
3 \Bigl( \frac{\partial \mathfrak{g}^{\m \a }}{\partial x_{\a}}+\mathfrak{g}^{\a \b} \G_{\a \b}^{\m} \Bigr)+
 \mathfrak{g}^{\m \a} \bigl(\G_{\a \b}^{\b}-\G_{\a \b}^{\b} \bigr)=0.
\ee

\be
\frac{\partial \mathfrak{g}^{\m \a }}{\partial x_{\a}}-
\frac{\partial \mathfrak{g}^{\a \n }}{\partial x_{\a}}=0.
\ee

If we further introduce the quantities $\mathfrak{g}_{\m \n}$ which are the normalized
subdeterminants of the $\mathfrak{g}^{\m \n}$ and thus fulfill the equations

\bd
\mathfrak{g}_{\m \a }  \mathfrak{g}^{\n \a } = \mathfrak{g}_{\a \m }  \mathfrak{g}^{\a \n } =  \d_{\m}^{\n}.
\ed
and if we now multiply (5) by $\mathfrak{g}_{\m \n}$, after pulling up one index
the result may be written as follows:
\be
2 \mathfrak{g}^{\m \a }  \Bigl( \frac{\partial \ {\rm lg} \sqrt{\mathfrak{g}}}{\partial x_{\a}}+
\G_{\a \b}^{\b} \Bigr) + \bigl( \G_{\a \b}^{\b}-\G_{\b \a}^{\b} \bigr)+
\d_{\m}^{\n} \Bigl( \frac{\partial \mathfrak{g}^{\b \a }}{\partial x_{\a}}+
 \mathfrak{g}^{\s \b} \G_{\s \b}^{\b} \Bigr)=0,
\ee while $\g$ denotes the determinant of $\mathfrak{g}_{\m \n}$. The
equations (6) and (8) we write in the form 
\be 
\mathfrak{f}^{\m}= \frac{1}{3}
\mathfrak{g}^{\m \a} \bigl( \G_{\a \b}^{\b}-\G_{\b \a}^{\b} \bigr)= -\Bigl(
\frac{\partial \mathfrak{g}^{\m \a }}{\partial x_{\a}}+\mathfrak{g}^{\a \b} \G_{\a
\b}^{\m} \Bigr)= -\mathfrak{g}^{\m \a }  \Bigl( \frac{\partial \ {\rm lg}
\sqrt{\mathfrak{g}}}{\partial x_{\a}}+ \G_{\a \b}^{\b} \Bigr),
\ee 
whereby $\mathfrak{f}^{\m}$ stands for a certain tensor density. It is
easy to prove that the system (5) is equivalent to the system 
\be
\frac{\partial \mathfrak{g}^{\m \n }}{\partial x_{\a}}+ \mathfrak{g}^{\b \n} \G_{\b
\a}^{\m}+ \mathfrak{g}^{\m \b} \G_{\a \b}^{\n}- \mathfrak{g}^{\m \n} \G_{\a \b}^{\b}+
\d_{\a}^{\n} \mathfrak{f}^{\m} =0 
\ee 
in conjunction with (7). By pulling
down the upper indices we obtain the relations \bd \mathfrak{g}_{\m \n} =
\frac{g_{\m \n}}{\sqrt{-g}}= g_{\m \n} \sqrt{-g}, 
\ed 
whereby$\mathfrak{g}_{\m \n}$ is a covariant tensor \bd \qquad \qquad \qquad \qquad
-\frac{\partial g_{\m \n }}{\partial x_{\a}}+ g_{\s \n} \G_{\m
\a}^{\s}+ g_{\m \s} \G_{\a \n}^{\s}+g_{\m \n} \p_{\a} +g_{\m \a}
\p_{\n}=0, \qquad \qquad \qquad \qquad \qquad \qquad (10a) 
\ed
whereby $\p_{\t}$ is a covariant vector. This system, together
with the two systems given above, 
\bd \qquad \qquad \qquad \qquad
\qquad \qquad \frac{\partial \mathfrak{g}^{\n \a }}{\partial
x_{\a}}-\frac{\partial \mathfrak{g}^{\a \n }}{\partial x_{\a}}=0 \qquad
\qquad \qquad \qquad \qquad \qquad \qquad \qquad (7) 
\ed and \bd
\qquad \qquad \qquad \qquad 0 = R_{\m \n }= - \frac{\partial
\G_{\m \n }^{\a}}{\partial x_{\a}}+ \G_{\m \b}^{\a} \G_{\a
\n}^{\b} + \frac{\partial \G_{\m \a }^{\a}}{\partial x_{\n}}-
\G_{\m \n}^{\a} \G_{\a \b}^{\b}, \qquad \qquad \qquad \qquad
\qquad \qquad (4) 
\ed are the result of the variational principle
in the most simple form. Looking at this result, it is remarkable
that the vector $\p_{\t}$ occurs besides the tensor $(\mathfrak{g}_{\m \n})$
and the quantities $\G_{\m \n}^{\a}$. To obtain consistency with
the known laws of gravitation end electricity,  we have to
interpret the symmetric part of $\mathfrak{g}_{\m \n}$ as metric tensor and
the skew-symmetric part as electromagnetic field, and we have to
assume the vanishing of $\p_{\t}$, which will be done in the
following. For a later analysis (e.g. the problem of the
electron), we will have to keep in mind that the Hamiltonian
principle does not indicate a vanishing $\p_{\t}$. Setting
$\p_{\t}$ to zero leads to an overdetermination of the field,
since we have $16+64+4$ algebraically independent differential
equations for $16+64$ variables.

\section{$\S~2.$ The pure gravitational field as special case}

Let the $g_{\m \n}$ be symmetric. The equations (7) are fulfilled
identically. By changing $\m$ to $\n$ in  (10a) and subtraction we
obtain in easily understandable notation \be \G_{\n, \m \a}
+\G_{\m, \a \n} -\G_{\m, \n \a} - \G_{\n, \a \m} =0. \ee If $\D$
is called the skew-symmetric part of $\G$ with respect to the last
two indices, (11) takes the form \bd \D_{\n, \m \a} +\D_{\m, \a
\n} =0 \ed or
\bd \qquad \qquad \qquad \qquad \D_{\n, \m \a} = \D_{\m, \n \a}. \qquad \qquad
\qquad \qquad \qquad \qquad \qquad \qquad (11a) \ed This symmetry
property of the first two indices contradicts the antisymmetry of
the last ones, as we learn from the series of equations 
\bd
\D_{\m, \n \a}  = - \D_{\m, \a \n} =-\D_{\a, \m \n} = \D_{\a, \n
\m} = \D_{\n, \a \m} =- \D_{\n, \m \a}. 
\ed This, in conjunction
with (11a), compels the vanishing of all $\D$. Therefore, the $\G$
are symmetric in the last two indices as in {\sc Riemann}ian geometry.
The equations (10a) can be resolved in a well-known manner, and
one obtains \be \frac{1}{2} g^{\a \b}  \Bigl(\frac{\partial g_{\m
\b }}{\partial x_{\n}}+ \frac{\partial g_{\n \b }}{\partial
x_{\m}}- \frac{\partial g_{\m \n }}{\partial x_{\b}} \Bigr). \ee
Equation (12), together with (4) is the well-known law of
gravitation. Had we presumed the symmetry of  the $g_{\m \n}$ at
the beginning, we would have arrived at (12) and (4) directly.
This seems to be the most simple and coherent derivation of the
gravitational equations for the vacuum to me. Therefore it should
be seen as a natural attempt to encompass the law of
electromagnetism by generalizing 
these considerations rightly. Had we not assumed the vanishing of
the $\p_{\t}$, we would have been unable to derive the known law
of the gravitational field in the above manner by assuming the
symmetry of the $g_{\m \n}$. Had we assumed the symmetry of  both
the $g_{\m \n}$ {\em and\/} the $\G_{\m \n}^{\a}$ instead, the
vanishing of $\p_{\a}$ would have been a consequence of (9) or
(10a) and (7); we would have obtained the law of the pure
gravitational field as well.

\section{$\S~3.$ Relations to {\sc Maxwell}'s theory}
If there is an electromagnetic field, that means the $\mathfrak{g}^{\m \n}$
or the $g_{\m \n}$ do contain a skew-symmetric part, we cannot
solve the eqns.~(10a) any more with respect to the $\G_{\m
\n}^{\a}$, which significantly complicates the clearness of the
whole system. We succeed in resolving the problem however, if we
restrict ourselves to the first approximation. We shall do this
and once again postulate the vanishing of $\p_{\t}$. Thus we start
with the ansatz \be g_{\m \n} = - \d_{\m \n} + \g_{\m \n}+ \p_{\m
\n}, \ee whereby the $\g_{\m \n}$ should be symmetric, and the
$\p_{\m \n}$ skew-symmetric, both should be infinitely small in
first order. We neglect quantities of second and higher orders.
Then the $\G_{\m \n}^{\a}$ are infinitely small in first order as
well.

Under these circumstances the system (10a) takes the more simple form
\bd
\qquad \qquad \qquad \qquad +\frac{\partial g^{\m \n }}{\partial x_{\a}}+\G_{\m \a}^{\n} + \G_{\a \n}^{\m}=0.
\qquad \qquad \qquad \qquad \qquad \qquad \qquad \qquad(10b)
\ed
After applying two cyclic permutations of the indices $\m$, $\n$ and $\a$
two further equations appear. Then, out of the three equations we may calculate
the $\G$ in a similar manner as in the symmetric case. One obtains
\be
-\G_{\m \n}^{\a} = \frac{1}{2}  \Bigl(\frac{\partial g_{\a \n }}{\partial x_{\m}}+
\frac{\partial g_{\m \a }}{\partial x_{\n}}-
\frac{\partial g_{\n \m }}{\partial x_{\a}} \Bigr).
\ee

Eqn.~(4) is reduced to the first and third term. If we put the expression
$\G_{\m \n}^{\a}$ from (14) therein, 
one obtains
\be
-\frac{\partial^2 g_{\n \m }}{\partial x_{\a}^2}+
\frac{\partial^2 g_{\a \m }}{\partial x_{\n} \partial x_{\a}}+
\frac{\partial^2 g_{\a \n }}{\partial x_{\m} \partial x_{\a}}-
\frac{\partial^2 g_{\a \a }}{\partial x_{\m} \partial x_{\n}}=0.
\ee
Before further consideration of (15), we develop the series from equation (7).
Firstly, out of (13) follows that the approximation we are interested in yields
\be
\mathfrak{g}^{\m \n} = - \d_{\m \n} - \g_{\m \n}+ \p_{\m \n},
\ee
Regarding this, (7) transforms to
\be
\frac{\partial \p_{\m \n }}{\partial x_{\n}} =0.
\ee
Now we put the expressions given by (13) into (15) and obtain with respect to (17)

\be
-\frac{\partial^2 \g_{\m \n }}{\partial x_{\a}^2}+
\frac{\partial^2 \g_{\m \a }}{\partial x_{\n} \partial x_{\a}}+
\frac{\partial^2 \g_{\n \a }}{\partial x_{\m} \partial x_{\a}}-
\frac{\partial^2 \g_{\a \a }}{\partial x_{\m} \partial x_{\n}}=0
\ee

\be
\frac{\partial^2 \p_{\m \n }}{\partial x_{\a}^2} =0.
\ee

The expressions (18), which may be simplified as usual by proper
choice of coordinates, are the same as in the absence of an
electromagnetic field. In the same manner, the equations (17) and
(19) for the electromagnetic field do not contain the quantities
$\g_{\m \n}$ which refer to the gravitational field. Thus both
fields are - in accordance with experience - independent in first
approximation. 

The equations  (17), (19) are nearly equivalent to
{\sc Maxwell}'s equations of empty space. (17) is one {\sc Maxwell}ian system.
The expressions 
\bd \frac{\partial \p_{\m \n }}{\partial x_{\a}}+
\frac{\partial \p_{\n \a }}{\partial x_{\m}}+ \frac{\partial
\p_{\a \m }}{\partial x_{\n}}, 
\ed which\footnote{This appears to
be a misprint. The first term should be squared.} according to
 {\sc Maxwell} should vanish, do not vanish necessarily due to (17)
and (19), but their divergences of the form 
\bd \frac{\partial
}{\partial x_{\a}} \Bigl( \frac{\partial \p_{\m \n }}{\partial
x_{\a}}+ \frac{\partial \p_{\n \a }}{\partial x_{\m}}+
\frac{\partial \p_{\a \m }}{\partial x_{\n}} \Bigr) 
\ed however
do. Thus (17) and (19) are substantially identical to {\sc Maxwell}'s
equations of empty space.

Concerning the attribution  of $\p_{\m \n}$ to the electric and magnetic
vectors ($\mathfrak{a}$ and $\mathfrak{h}$) I would like to make a comment
 that claims validity independently from the theory
presented here. According to classical mechanics that uses
central forces to every sequence of motion $V$ there is an inverse motion $\bar V$,
that passes the same configurations by taking an inverse 
succession. This inverse motion $\bar V$ is formally obtained from $V$
by substituting
\bea  \nonumber
x^{'} =x \\ \nonumber
y^{'} =y \\ \nonumber
z^{'} =z \\ \nonumber
t^{'} =-t  \nonumber
\eea
in the latter one.

We observe a similar behavior, according to the general theory of
relativity, in the case of a pure gravitational field. To achieve
the solution $\bar V$ out of $V$, one has to substitute $t' = -t$
into all field functions and to change the sign of the field
components $g_{14}$, $g_{24}$, $g_{34}$ and the energy components
$T_{14}$, $T_{24}$, $T_{34}$. This is basically the same procedure
as applying the above transformation to the primary motion $V$.
The change of signs in $g_{14}$, $g_{24}$, $g_{34}$ and in
$T_{14}$, $T_{24}$, $T_{34}$ is an intrinsic consequence of the
transformation law for tensors.

This generation of the inverse motion by transformation of the
time coordinate ($t' = -t$) should be regarded as a general law
that claims validity for electromagnetic  processes as well.
There, an inversion of the process changes the sign of the
magnetic components, but not those of the electric ones. Therefore
one should have to assign the components $\p_{23}$, $\p_{31}$,
$\p_{12}$ to the electric field and $\p_{14}$, $\p_{24}$,
$\p_{34}$ to the magnetic field. We have to give up
the inverse assignment which was in use as yet. 
It was preferred so far, since it seems more comfortable  to express the density
of a current by a vector rather than by a skew-symmetric tensor of third rank.

Thus in the theory outlined here, (7) respectively (17) is the expression for the law
of magnetoelectric induction. In accordance, at the r.h.s. of the equation
there is no term that could be interpreted as density of the electric current.

The next issue is, if the theory developed here renders the
existence of singularity-free, centrally 
 symmetric electric masses comprehensible. I started to tackle this problem together with
Mr. J.~{\sc Grommer}, who was at my disposal ceaselessly  for all
calculations while analyzing the general theory of relativity in
the last years. At this point I would like to express my best
thanks to him and to the `International educational board' which
has rendered possible the continuing collaboration with
Mr.~{\sc Grommer}.

\newpage
\setcounter{section}{0}
\setcounter{equation}{0}
\setcounter{page}{1}

\v
\begin{center}
\huge{Riemannian Geometry with Maintaining the Notion \\
of Distant Parallelism\\}
\v
\Large
Albert~Einstein \\

\normalsize
translation by A.~Unzicker and T.~Case\\
\v
\large{Session Report of the Prussian Academy of Sciences, pp.~217-221\\
June 7th, 1928}
\end{center}
\v

Riemannian Geometry has led to a physical description of the
gravitational field in the theory of general relativity, but
it did not provide concepts that can be attributed to the
electromagnetic field. Therefore, theoreticians aim to
find natural generalizations or extensions of riemannian
geometry that are richer in concepts, hoping to arrive at
a logical construction that unifies all physical field
concepts under one single leading point. Such endeavors
brought me to a theory which should be communicated even
without attempting any  physical interpretation,
because it can claim a certain interest just
because of the naturality of the concepts introduced therein.

Riemannian geometry is characterized by an Euclidean
metric in an infinitesimal neighborhood 
of any point $P$. Furthermore, the absolute values of the line
elements which  belong to the neighborhood of two points $P$ and
$Q$ of finite distance can be compared. However, the notion of
parallelism of such line elements is missing; a concept of
direction does not exist for the finite case. The theory outlined
in the following is characterized by introducing - beyond the
Riemannian metric- the concept of `direction', `equality of
directions' or
 `parallelism' for finite distances. Therefore, new
invariants and tensors will arise besides those known
in Riemannian geometry.

\section{$\S~1.$ $n$-bein field and metric}

Given an arbitrary point $P$ of the $n$-dimensional continuum,
let's imagine an orthogonal $n$-bein of $n$ unit vectors that
represents a local coordinate system. $A_a$ are the components of
a line element or another vector with respect to this local system
($n$-bein). Besides that, we introduce a Gaussian
coordinate system of the $x^{\n}$ for describing a finite domain. 
Let $A^{\n}$ be the components of a vector (A) with respect to the
latter, and $h_{\ a}^{\n}$ the $\n$-components of the unit vectors
forming the $n$-bein. Then, we have\footnote{We assign Greek
letters to the coordinate indices and Latin ones to the bein
indices.} \be A^{\n}= h_{\ a}^{\n} A_{a}.     \dots \ee
One obtains the inversion 
of (1) by calling $h_{\n a}^{\n}$ the normalized subdeterminants
of the $h_{\ a}^{\n}$, \bd A_{a}= h_{\m a} A^{\m} \dots .\qquad
\qquad (1a) \ed \setcounter{equation}{1} Since the infinitesimal
sets are Euclidean, \be A^{2} = \sum{A^{2}_a} = h_{\m a} \ h_{\n
a} A^{\m} A^{\n} \dots \ee holds for the modulus $A$
 of the vector (A).

Therefore, the components of the metric tensor appear in the form
\be
g_{\m \n} = h_{\m a}\  h_{\n a}, \dots
\ee
whereby the sum has to be taken over $a$. For a fixed $a$, the
$h_{a}^{\m}$ are the components of a contravariant vector.
Furthermore, the following relations hold:
\bea
h_{\m a} \ h_{a}^{\n}=\d_{\m}^{\n}  \dots    \\
h_{\m a} \ h_{b}^{\m}=\d_{a b}, \dots \eea with $\d=1$ if the
indices are equal, and $\d=0$, if not. The correctness of (4) and
(5) follows from the above definition of the $h_{\m a}$ as the
normalized subdeterminants of the $h_{a}^{\m}$. The vector
property of $h_{\m a}$ follows conveniently from the fact that the
l.h.s. and therefore, the r.h.s. of (1a) as well, are invariant
for any coordinate transformation and for any choice of the vector
(A). The $n$-bein field is determined by $n^2$ functions
$h_{a}^{\m}$, whereas the Riemannian metric is determined just by
$\frac{n(n+1)}{2}$ quantities. According to (3), the metric is
determined by the $n$-bein field but not vice versa.

\section{$\S~2.$ Teleparallelism and rotation invariance}

By postulating the existence of the $n$-bein field (in every
point) one expresses implicitly the existence of a Riemannian
metric and distant parallelism. ($A$) and ($B$) being  two vectors
in the points $P$ and $Q$ which have the same local coordinates
with respect to their $n$-beins (that means $A_a = B_a$), then
have to be regarded as equal (because of (2)) and as 'parallel'.

If we take the metric and the teleparallelism as the essential, i.e. the
objective meaningful things, then we realize that the $n$-bein field is
not yet fully determined by these settings. Yet metric and teleparallelism
remain intact, if we substitute the $n$-beins of all points of the continuum
with such $n$-beins that were derived out of the original ones by the  rotation
stated above. We denote this substitutability of the $n$-bein field as
rotational invariance and establish: 
Only those mathematical relations that are rotational invariant can claim a real
meaning.

Thus by keeping the coordinate system fixed, and a given metric
and parallel connection, the $h_a^{^\m}$ are not yet fully
determined; there is a possible substitution which corresponds to
the
rotation invariance 
\be A_a^* = d_{a\ m} A_m ..., \ee whereby $d_{a \ m}$ is chosen
orthogonal and independent of the coordinates. ($A_a$) is an
arbitrary vector with respect to the local system, ($A^*_a$) the
same vector with respect to the rotated local system. According to
(1a), and using (6), it follows \bd h_{\m a}^{\ *} A^{\m} = d_{a
m} h_{\m m} A^{\m}     \nonumber \ed or \bd h_{\m a}^{*} =d_{a m}
h_{\m m},  \dots   \qquad \qquad (6a) \ed whereby \bd d_{a m} \
d_{b m} = d_{m a} \ d_{m b}= \d_{a  b} , \dots  \qquad  \qquad
(6b) \ed \bd \frac{\partial d_{a m}}{\partial x^{\n}} =0. \dots
\qquad  \qquad (6c) \ed
Now the postulate of rotation invariance 
tells us that among the relations in  which the quantities $h$
appear, only those may be seen as meaningful, which are transformed
into $h^*$ of equal form, if $h^*$ is introduced by eqns. (6). In other
words: $n$-bein fields which are related by locally equal
rotations are equivalent.

The rule of infinitesimal parallel transport of a vector from
point ($x^{\n}$) to a neighboring point ($x^{\n} + d x^{\n}$) is
obviously characterized by \be d \ A_a =0 \dots ,
\ee that means by
the equation
\bd 
0= d(h_{\m a} A^{\m}) = \frac{\partial h_{\m a}}{\partial x^{\s}}
A^{\m} d x^{\s} + h_{\m a} d A^{\m} =0 \ed Mulitplicated by
$h_a^{\n}$ this equation becomes considering (5) \bd d A^{\n} = -
\D_{\m \s}^{\n}  A^{\m} d x^{\s} \qquad \qquad (7a) \ed with \bd
\D_{\m \s}^{\n} = h_{a}^{\n} \frac{\partial h_{\m a}}{\partial
x^{\s}}. \ed This law of parallel transport is rotation invariant
and not symmetric with respect to the lower indices of the
quantities $\D_{\m \s}^{\n} $. If one transports  the vector (A)
now according to this law along a closed path, the vector remains
unaltered; this means, that the Riemannian tensor \bd R_{k, l
m}^{i} = -\frac{\partial \D_{k l}^{i} }{\partial x^{m}}+
\frac{\partial \D_{k m}^{i} }{\partial x^{l}}+ \D_{\a l}^{i} \D_{k
m}^{\a} -\D_{\a m}^{i} \D_{k l}^{\a} \ed built from the connection
coefficients vanishes according to (7a), which can be verified
easily. Besides this law of parallel transport there is that
(nonintegrable) symmetric transport law due to the Riemannian
metric (2) and (3). As is generally known, it is given by the
equations \bea \bar d A^{\n} = -\G_{\m \t}^{\n} A^{\m} d x^{\t} \\
\nonumber \G_{\m \t}^{\n}  = \frac{1}{2} g^{\n \a} \bigl(
\frac{\partial g_{\m \a}}{\partial x^{\t}}+ \frac{\partial g_{\t
\a}}{\partial x^{\m}}- \frac{\partial g_{\m \t}}{\partial x^{\a}}
\bigr). \eea According to (3), the $\G_{\m \t}^{\n} $
 are expressed by the quantities $h$ of the $n$-bein fields.
Thereby one has to keep in mind that
\be
g^{\m \n} = h_{\a}^{\m} h_{\a}^{\n}. ...
\ee
Because of this setting and due to (4) and (5) the equations
\bd
g^{\m \l} g_{\n \l} = \d_{\n}^{\m}
\ed
are fulfilled which define the $g^{\m \l}$ calculated from the $g_{\m \l}$.
This transport law based on metric only is obviously rotation
invariant in the above sense.

\section{$\S~3.$ Invariants and covariants}

On the manifold we are considering, besides the tensors
and invariants of {\sc Riemann}-geometry which
contain the quantities $h$ only in the combination
(3), other tensors and invariants exist,
among which we will have a look at the simplest
ones only.

If one starts with a vector $(A^{\n})$ in the point $x^{\n}$, with
the shifts $d$ and $\bar d$, the two vectors \bd A^{\n} + d A^{\n}
\ed and \bd A^{\n} + \bar d A^{\n} \ed are produced in the
neighboring point $(x+ d x^{\n})$. Thus the difference \bd d
A^{\n} - \bar d A^{\n} = (\G_{\a \b}^{\n} -\D_{\a \b}^{\n}) A^{\a}
d x^{\b} \ed has vector character as well. Therefore, \bd (\G_{\a
\b}^{\n} -\D_{\a \b}^{\n}) \ed is a tensor, and also its
skewsymmetric part \be \frac{1}{2} (\D_{\a \b}^{\n} -\D_{\b
\a}^{\n}) = \L_{\b \a}^{\n} \dots \ee The fundamental meaning of
this tensor in the theory developed here results from the
following: If this tensor vanishes, then the continuum is
Euclidean. Namely, if \bd 0= 2 \L_{\a \b}^{\n} = h_a^{\n}
(\frac{\partial h_{\a a}}{\partial x^{\b}}- \frac{\partial h_{\b
a}}{\partial x^{\a}}), \ed
 holds, then by multiplication with $h_{\n b}$
follows
\bd
0= \frac{\partial h_{\a b}}{\partial x^{\b}}-
\frac{\partial h_{\b b}}{\partial x^{\a}}.
\ed
However, one may assume
\bd
h_{a b} = \frac{\partial \psi_b}{\partial x^{\a}}.
\ed
Therefore the  field is derivable from $n$ scalars
$\psi_b$. We now choose  the coordinates according
to the equation
\bd
\psi_b= x^b
\ed
Then, due to (7a) all the $\D_{\b \a}^{\n}$ vanish, and the
$h_{\m a}$ and the $g_{\m \n}$ are constant.---

Since the tensor\footnote{tr. note:
this is called torsion tensor in the literature.} $\L_{\b \a}^{\n}$ is formally
the simplest one admitted by our theory, this tensor
 shall be used as a starting point for characterizing such
a continuum, and not the more complicated Riemannian
curvature tensor. The most simple quantities which
come in mind are the vector
\bd
\L_{\m \a}^{\a}
\ed
and the invariants
\bd
g^{\m \n} \L_{\m \b}^{\a} \L_{\n \a}^{\b} \ \ \mbox{and} \ \
g_{\m \n} g^{\a \s} g^{\b \t} \L_{\a \b}^{\m} \L_{\s \t}^{\n}
\ed
From one of the latter ones (actually,
from a linear combination of it), after multiplication with
the invariant volume element
\bd
h \ d \t,
\ed
(whereby $h$ means the determinant $\mid h_{\m \a} \mid$,
$d \t$ the product $ d x_1... d x_n$), an invariant integral $J$,
may be built.
The setting
\bd
\d J = 0
\ed
then provides  16 differential equations for the
16 quantities $h_{\m \a}$.

If laws with relevance
to physics can be derived from this, shall be investigated later.---
It clarifies things, to compare 
{\sc Weyl's\/}
modification of the {\sc Riemannian\/} theory
to the one presented here:
\begin{quote}
{\sc Weyl\/}: no comparison at a distance, neither
of the absolute values, nor of directions of vectors.\\
{\sc Riemann\/}: comparison at a distance for
absolute values of vectors, but not of directions of vectors.\\
{\sc Present theory\/}: comparison of both absolute values and
directions of vectors at a distance.\footnote{tr. note: This is
the origin of the name {\em distant parallelism\/} as a synonym
for {\em absolute  parallelism\/} or
 {\em teleparallelism\/}, in German
{\em Fernparallelismus\/}.}
\end{quote}

\newpage
\setcounter{section}{0}
\setcounter{equation}{0}
\setcounter{page}{1}

\v
\begin{center}
\huge{New Possibility for  a Unified Field Theory\\
of Gravitation and Electricity\\}
\v
\Large
Albert~Einstein \\

\normalsize
translation by A.~Unzicker and T.~Case\\
\v
\large{Session Report of the Prussian Academy of Sciences, pp.~224-227\\
June 14th, 1928}
\end{center}
\v

Some days ago I explained in a short note in these
reports, how by using a $n$-bein field
a geometric theory  can be constructed that is
based on the notion of a Riemann-metric and distant
parallelism. I left open the question 
if this theory could serve for describing 
physical phenomena. 
In the meantime I discovered that this theory - at least in
first approximation -- yields the field equations
of gravitation and electromagnetism in a very
simple and natural manner. Thus it seems possible
that this theory will substitute the theory
of general relativity in its original form.

The introduction of this theory has as a consequence the existence
of a straight line, that means a line of which all elements are
parallel to each other.\footnote{tr. note: such lines are nowadays
called autoparallels.} Naturally, such a line is not identical
with a geodesic. Furthermore, contrarily to the actual theory of
relativity, the notion of relative rest of two mass points exists
(parallelism of two line elements that belong to two different
world lines).

In order to apply the general theory in the implemented 
form to the field theory, one has to set the following
conventions:
\begin{quote}
1. The dimension is 4 ($n=4$).\\
2. The fourth local component $A_a$ (a=4) of a vector is purely
imaginary, and so are the components of the 4th bein of the
4-bein,\footnote{tr.note: 4-bein (tetrad, from German `bein' =
leg) has become a common expression in differential geometry)} and
also the quantities $h_4^{\n}$ and $h_{\n 4}$\footnote{Instead of
this one could define the square of the length of the local vector
$A_1^2+A_2^2+A_3^2-A_4^2$ and introduce Lorentz-transformations
instead of rotations of the local $n$-bein. In that case all the
$h$ would become real, but one would loose the direct connection
to the formulation of the general theory.}. Of course, all the
coefficients of $\g_{\m \n}$  (= $h_{\m a} h_{\n a}$) become then
real. Thus, we choose the square of the modulus 
of a timelike vector to be negative.
\end{quote}

\section{$\S~1.$ The assumed basic field law}
For the  variation of the field potentials
$h_{\m a}$ (or $h_{\a}^{\m}$) to vanish on the
boundary of a domain the variation of the Hamiltonian
integral should vanish:
\be
\d \{ \int {\cal H} d \t \} = 0,
\ee
\bd
\qquad   \qquad  {\cal H} = h \ g^{\m \n} \ \L_{\m \b}^{\ \a} \
\L_{\n \a}^{\ \b}, \qquad   \qquad \qquad   \qquad   \qquad  (1a)
\ed
with the quantities $h (= \mid h_{\m \a} \mid)$,
$g^{\m \n}$, $\L_{\m \n}^{\ \a} $ defined in the
eqns.~(9), (10) loc.it.\footnote{tr. note: this
refers to the session report of June 7th, 1928.}

The field $h$ should describe both the electrical
and gravitational field. A `pure gravitational'
field means 
that in addition to eqn.~(1) being satisfied
the quantities
\be
\p_{\m} =\L_{\m \a}^{\ \a}
\ee
vanish, which is a covariant and rotation invariant
restriction.\footnote{There is still a certain
ambiguity in interpreting, because one could
characterize the gravitational field by the
vanishing of
$\frac{\partial \p_{\m}}{\partial x_{\n}}-
\frac{\partial \p_{\n}}{\partial x_{\m}}$
as well.}

\section{$\S~2.$ The field law in first approximation}

If the manifold is the {\sc Minkowski\/}-world of the theory of
special relativity, one may choose the coordinate system in a way
that $h_{11} = h_{22} = h_{33} =1, h_{44} =  j \ (=\sqrt{-1})$
holds and the other $h_{\m \a}$ vanish. This system of values for
$h_{\m \a}$ is a little inconvenient for calculations. Therefore
in this paragraph for calculations we prefer to assume the
$x_4$-coordinate to be purely imaginary; then, the {\sc
Minkowski\/}-world (absence of any field for a suitable choice of
coordinates) can be described by \be h_{\m \a} = \d_{\m \a} \dots
\ee The case of infinitely weak fields can be described
purposively by \be
 h_{\m \a} = \d_{\m \a} + k_{\m \a}, \dots
\ee
whereby the $ k_{\m \a}$ are small values of first order.
While neglecting quantities of third and higher order
one has to replace (1a) with respect to (10) and (7a)
loc. it. by
\bd
\qquad \qquad {\cal H} = \frac{1}{4} \bigl(
\frac{\partial k_{\m \a}}{\partial x_{\b}}-
\frac{\partial k_{\b \a}}{\partial x_{\m}}
\bigr)
\bigl(
\frac{\partial k_{\m \b}}{\partial x_{\a}}-
\frac{\partial k_{\a \b}}{\partial x_{\m}}
\bigr). \qquad \qquad \qquad \qquad \qquad (1b)
\ed
By performing the variation one obtains the field
equations valid in first approximation
\be
\frac{\partial^2 k_{\b \a}}{\partial x^2_{\m}}-
\frac{\partial^2 k_{\m \a}}{\partial x_{\m} \partial x_{\b}}+
\frac{\partial^2 k_{\a \m}}{\partial x_{\m} \partial x_{\b}}-
\frac{\partial^2 k_{\b \m}}{\partial x_{\m} \partial x_{\a}} =0.  \dots
\ee
This are 16 equations\footnote{Naturally, between the
 field equations there exist four identities due to the
general covariance. In the first approximation treated here this
is expressed by the fact that the divergence taken with respect to
the index $a$ of the l.h.s. of (5) vanishes identically.} for the
16 quantities $k_{\a \b}$. Our task is now to see if this system
of equations contains the known laws of gravitational and the
electromagnetical field. For this purpose we introduce in (5) the
$g_{\a \b}$ and the $\p_{\a}$ instead of the $k_{\a \b}$.
We have to define 
\bd
g_{\a \b} = h_{\a a} h_{\b a} =
(\d_{\a a} + k_{\a a}) (\d_{\b a} + k_{\b a})
\ed
or in first order
\be
g_{\a \b}-\d_{\a \b} = \overline{g_{\a \b}} =
k_{\a \b} + k_{\b \a}. \dots
\ee
From (2) one obtains further the quantities
of first order, precisely
\bd
 2 \ \p_{\a} =
\frac{\partial k_{\a \m}}{\partial x_{\m}}-
\frac{\partial k_{\m \m}}{\partial x_{\a}}. \dots
\qquad  \qquad \qquad \qquad (2a)
\ed
By exchanging $\a$ and $\b$ in (5) and adding the
thus obtained structuring of (5)  at first one gets
\bd
\frac{\partial^2 \overline{g_{\a \b}}}{\partial x^2_{\m}}-
\frac{\partial^2 k_{\m \a}}{\partial x_{\m} \partial x_{\b}}-
\frac{\partial^2 k_{\m \b}}{\partial x_{\m} \partial x_{\a}} =0.
\ed
If to this equation the two equations
\bea \nonumber
- \frac{\partial^2 k_{\a \m}}{\partial x_{\m} \partial x_{\b}} +
\frac{\partial^2 k_{\m \m}}{\partial x_{\a} \partial x_{\b}} =
-2 \frac{\partial \p_{\a}}{\partial x_{\b}} \\ \nonumber
- \frac{\partial^2 k_{\b \m}}{\partial x_{\m} \partial x_{\a}} +
\frac{\partial^2 k_{\m \m}}{\partial x_{\a} \partial x_{\b}} =
-2 \frac{\partial \p_{\b}}{\partial x_{\a}}, \nonumber
\eea
are added, following from (2a), one obtains according to (6)
\be
\frac{1}{2} \bigl(
- \frac{\partial^2 \overline{g_{\a \b}}}{\partial x^2_{\m}} +
\frac{\partial^2 \overline{g_{\m \a}}}{\partial x_{\m} \partial x_{\b}} +
\frac{\partial^2 \overline{g_{\m \b}}}{\partial x_{\m} \partial x_{\a}} -
\frac{\partial^2 \overline{g_{\m \m}}}{\partial x_{\a} \partial x_{\b}}+
\bigr)
= \frac{\partial \p_{\a}}{\partial x_{\b}} +
\frac{\partial \p_{\b}}{\partial x_{\a}}. \dots
\ee
The case of the absence of an electromagnetic field is
characterized by the vanishing of $\p_{\m}$. In this case
(7) is in first order equivalent to the equation
\bd
R_{\a \b} =0
\ed
used as yet in the theory of general relativity
($R_{\a \b}$ = contracted Riemann tensor). {\em
With the help of this it is proved that our new theory yields
the law of a pure gravitational field in first approximation correctly\/}.

By differentiation of (2a) by $x_{\a}$, one gets  the equation
\be
\frac{\partial \p_{\a}}{\partial x_{\a}} =0.
\ee
according to (5) and contraction over $\a$ and $\b$.
Taking into account that the l.h.s. $L_{\a \b}$ of (7)
fulfills the identity
\bd
\frac{\partial}{\partial x_{\b}} (L_{\a \b} - \frac{1}{2}
\d_{\a \b} L_{\s \s}) =0,
\ed
from  (7) follows
\bd
\frac{\partial^2 \p_{\a}}{\partial^2 x_{\b}} +
\frac{\partial^2 \p_{\b}}{\partial x_{\a} \partial x_{\b}}-
\frac{\partial}{\partial x_{\a}}
\bigl(\frac{\partial \p_{\s}}{\partial x_{\s}} \bigr) =0
\ed
or
\be
\frac{\partial^2 \p_{\a}}{\partial^2 x_{\b}} =0. \dots
\ee
The equations (8) and (9) are, as it is well known,
equivalent to {\sc Maxwell}'s equations for empty space.
{\em The new theory thus also yields {\sc Maxwell}'s equations\/}
in first approximation.

According to this theory, the separation of
the gravitational and electromagnetic field
seems arbitrary however. Furthermore, it is clear
that the eqns.~(5) state more than the eqns.~(7), (8)
and (9) together. After all it is remarkable that
the electric field does not enter the field equations
quadratically.
\vspace{1.0cm}

{\bf Note added in proof.} One obtains very similar results by
starting with the Hamilton function \bd {\cal H} = h \ g_{\m \n}
g^{\a \s} g^{\b \t} \ \L_{\a \b}^{\ \m} \L_{\s \t}^{\ \n}. \ed
Thus for the time being there remains a certain insecurity
regarding the choice of ${\cal H}$. 

\newpage
\setcounter{section}{0}
\setcounter{equation}{0}
\setcounter{page}{1}

\v
\begin{center}
\huge{Unified Field Theory Based on Riemannian Metrics
and Distant Parallelism\\}
\v
\Large
Albert~Einstein \\

\normalsize
translation by A.~Unzicker and T.~Case\\
\v
\large{Math. Annal. 102 (1930), pp.~685-697}
\end{center}
\v

In the present work I would like to describe a theory I have been
working on for a year; it will be exposed in a manner that it can
be understood comfortably by  everyone who is familiar with the
theory of general relativity. The following exposure is necessary,
because due to coherences and improvements found in the meantime
reading the earlier work would be a useless loss of time. The
topic is presented in a way that seems most serviceable for
comfortable access. I learned, especially with the help of Mr.
Weitzenb\"ock and Mr. Cartan, that the dealing with the continua
we are talking about is not new. Mr. Cartan kindly wrote an essay
about the history of the relevant mathematical topic in order to
complete my paper; it is printed right after this paper in the
same review. I would also like to thank Mr. Cartan heartily at
this point for his valuable contribution. The most important and
undisputable new result of the present work is the finding of the
most simple field laws that can be applied to a Riemannian
manifold with distant parallelism. I am only going to discuss
their physical meaning briefly.

\section{$\S~1.$ The structure of the continuum}
Since the number of dimensions has no impact on the following considerations,
we suppose a  $n$-dimensional continuum. To take into account the facts
of  metrics  and  gravitation we assume the existence of
a Riemann-metric. In nature there also exist electromagnetic fields,
which cannot be described by Riemannian metrics.
This arouses the question:
How can we complement our Riemannian spaces in a natural,  logical
way with an additional  structure, so that the whole thing has a uniform character~?\\
The continuum is (pseudo-)Euclidean in the vicinity of every point
$P$. In every point there exists a local coordinate system of
geodesics (i.e. an orthogonal $n$-bein), in relation to which the
theorem of Pythagoras  is valid. The orientation of these
$n$-beins is not important in a Riemannian manifold. We would now
like to assume that these elementary Euclidean spaces are
governed by still another direction law. We are also going to
assume,  that it makes sense to speak of a  parallel orientation
of all $n$-beins together, applying  this to space structure like
in Euclidean geometry (which would be senseless in a  space with
metrical structure {\em only\/}).

In the following we are going to think of the  orthogonal
$n$-beins as being always in parallel  orientation. The in its
self arbitrary orientation of the local $n$-bein in one point  $P$
then determines the orientation of the  local $n$-beins  in all
points of the continuum uniquely. Our task now is to set up the
most simple restrictive laws which can be applied to such a
continuum. Doing so, we hope to derive the general laws of nature,
as the previous theory of general relativity tried this for
gravitation by applying a purely metrical space structure.

\section{$\S~2.$ Mathematical description of the space structure}

The local $n$-bein consists of $n$  orthogonal unit vectors with
components   $h_s^{\ \nu}$ with respect to any Gaussian coordinate
system. Here as always a lower Latin index indicates the
affiliation to a certain bein of the $n$-beins,  a Greek index -
due to its  upper or lower position - the covariant or
contravariant  transformation character of the relevant entity
with respect to a change of the Gaussian coordinate system. The
general transformation property of the $h_s^{\ \nu}$ is the
following. If all local systems or $n$-beins are twisted in the
same manner, which is a correct operation of course, and a new
Gaussian coordinate system is introduced at the same time,
 the following transformation law then exists in-between the
new and old  $h_s^{\ \nu}$
\begin{equation}
h_{s}^{\ \nu'}=\alpha_{st} \frac{\partial x^{\nu'}}{\partial x^{\alpha}} h_{t}^{\ \alpha},
\label{e1}
\end{equation}

\ni
whereas the constant coefficients $\alpha_{s t}$ form an orthogonal
system:

\begin{equation}
\alpha_{sa} \alpha_{sb}=\alpha_{as} \alpha_{bs}=\delta_{ab}=
\left\{  \begin{array}{ll}
1, & \mbox{if $a=b$}\\
0, & \mbox{if $a \not= b$}
\end{array} \right.
 \label{e2}
\end{equation}

Without problems the transformation law   (1) can be generalized
onto objects which components bear an arbitrary number of
 local indices  and coordinate indices. We call such objects
tensors. Out of this the algebraic  laws of  tensors (addition,
multiplication, contraction by Latin and Greek indices) follow
immediately.\\
We call  $h_s^{\ \nu}$ the components of the fundamental tensor. If a
 vector in the  local system has components $A_s$, and the coordinates
 $A^{\nu}$ with respect to the  Gaussian system, it follows out of
the meaning of the  $h_s^{\ \nu}$:
\begin{equation}
A^{\nu}=h_s^{\ \nu} A_s
\label{e3}
\end{equation}

\ni
or -- resolved  with respect to the $A_s$--
\begin{equation}
A_{s}=h_{s \nu} A^{\nu}
\label{e4}
\end{equation}

The tensorial character of the  normalized subdeterminants $h_{s \nu}$
of the
$h_s^{\ \nu}$ follows out of (4). $h_{s \nu}$ are the
covariant components of the fundamental tensor. Between $h_{s \nu}$ and
$h_s^{\ \nu}$ there are the relations

\begin{equation}
h_{s \mu} h_{s}^{\ \nu}=\delta_{\mu}^{\ \nu}=
\left\{  \begin{array}{ll}
1, & \mbox{if $\mu=\nu$}\\
0, & \mbox{if $\mu \not= \nu$}
\end{array} \right.
\label{e5}
\end{equation}

\begin{equation}
h_{s \mu} h_{t}^{\ \mu}=\delta_{st}
\label{e6}
\end{equation}

\ni
Due to the orthogonality of the local system we obtain the absolute value
 of the vector
\addtocounter{equation}{-1}
\begin{equation}
A^2 = A_s^2 =h_{s \mu}\  h_{s \nu} \ A^{\mu}\  A^{\nu} =g_{\mu \nu} \ A^{\mu} \ A^{\nu};
\label{e6a}
\end{equation}

\ni
Therefore,

\begin{equation}
g_{\mu \nu} = h_{s \mu}\  h_{s \nu}
\label{e7}
\end{equation}

are the  coefficients of the metric.\\
The  fundamental tensor allows (cfr. (3) and (4)) to
transform local indices into coordinate indices and vice versa (by
multiplication and contraction), so that it comes down to pure convention,
with which type of  tensors  one likes to operate.\\
Obviously the following relations hold:

\begin{displaymath}
A_{\nu}=h_{s \nu} A_s,   \qquad \qquad \qquad (3a)
\end{displaymath}

\begin{displaymath}
A_{s}=h_s^{\ \nu} A_{\nu}.     \qquad \qquad \qquad (4a)
\end{displaymath}

Furthermore, we have the relation of  determinants

\begin{equation}
g=\vert g_{\sigma \tau} \vert= {\vert h_{\alpha \sigma} \vert}^2= h^2
\label{e8}
\end{equation}

Therefore, the invariant of the volume element $\sqrt{g d \tau}$ takes
the form $h d \tau$.
To take into account the particular properties of time, it is most
comfortable to set the $x^4$-coordinate (both local and general)
of our 4-dimensional  space-time continuum
purely imaginary and also all tensor components with an odd
number of indices 4.

\section{$\S~3.$ Differential relations}

Now we denote   $\delta$ the change of the  components of a vector
or tensor during a  'parallel displacement' in the sense of
Levi-Civita during the transition to a infinitely neighboring
point of the continuum; now it follows out of the above

\begin{equation}
0=\delta A_s =\delta (h_{s \alpha} A^{\alpha})=\delta (h_s^{\ \alpha} A_{\alpha})
\label{e9}
\end{equation}

\ni
Resolving the brackets yields

\begin{eqnarray}
h_{s \alpha}\  \delta A^{\alpha} +A^{\alpha}\  h_{s \alpha, \beta}\ \delta x^{\beta}=0,
\nonumber\\
h_{s}^{\ \alpha}\  \delta A_{\alpha} +A_{\alpha}\  h_{s \ ,\beta}^{\ \alpha}\ \delta x^{\beta}=0,
\nonumber
\end{eqnarray}

\ni
whereas the colon indicates ordinary differentiation by
$x^{\beta}$. Resolving of the  equation yields

\begin{equation}
\delta A^{\sigma} =- A^{\alpha} \Delta^{\ \sigma}_{\alpha \ \beta}\ \delta x^{\beta},
\label{e10}
\end{equation}

\begin{equation}
\delta A_{\sigma} = A_{\alpha} \Delta_{\sigma \ \beta}^{\ \alpha}\ \delta x^{\beta},
\label{e11}
\end{equation}

\ni
whereby we set

\begin{equation}
\Delta_{\alpha \ \beta}^{\ \sigma} =h_{s}^{\ \sigma} \ h_{s \alpha, \beta}=
-h_{s \alpha}\  h_{s \ , \beta}^{\ \sigma}
\footnote{tr. note: the connection $\Delta$ is nowadays usually denoted as
$\Gamma$. Cfr. Schouten, Ricci Calculus (Springer, 1954), chap. III (1.2)}\label{e12}
\end{equation}

(The last conversion is based on(5)).\\
This law of parallel displacement is --~contrarily to Riemannian
geometry~-- in general not symmetric. If it is, we have Euclidean
geometry,  because\\

\begin{displaymath}
\Delta_{\alpha\ \beta}^{\ \sigma} -\Delta_{\beta\ \alpha}^{\ \sigma}=0
\end{displaymath}
or

\begin{displaymath}
h_{s \alpha,\beta}-h_{s \beta,\alpha}=0.
\end{displaymath}
But then

\begin{displaymath}
h_{s \alpha}=\frac{\partial \psi_s}{\partial x_{\alpha}}
\end{displaymath}
holds. If one chooses the $\psi_{s}$ as new variables $x'_s$, we obtain

\begin{equation}
h_{s \alpha}=\delta_{s \alpha},
\label{e13}
\end{equation}

\ni
proving the statement.\\

{\bf Covariant differentiation.} \rm
The local components of a vector are  invariant with respect to any
coordinate transformation. Out of this follows immediately the tensorial
character of the differential quotient

\begin{equation}
A_{s, \alpha}.
\label{e14}
\end{equation}

\ni
Because of  (4a) this can be replaced by

\begin{displaymath}
(h_s^{\ \sigma} A_{\ \sigma})_{,\alpha},
\end{displaymath}

\ni
and the tensorial character of

\begin{displaymath}
h_s^{\ \sigma} A_{\sigma, \alpha}+ A_{\sigma}\  h_{s \ ,\alpha}^{\ \sigma},
\end{displaymath}
follows. Equally (after multiplication with $h_{s \tau}$) the tensorial
character of
\begin{displaymath}
A_{\tau,\alpha} +  A_{\sigma}\  h_{s \ , \alpha}^{\ \sigma}\  h_{s \tau}
\end{displaymath}

\ni
and of

\begin{displaymath}
A_{\tau,\alpha} -  A_{\sigma}\  h_{s}^{\ \sigma} h_{s \tau,\alpha}
\end{displaymath}
and (see (16)) of

\begin{displaymath}
A_{\tau,\alpha} -  A_{\sigma} \Delta_{\tau \ \alpha}^{\ \sigma}.
\footnote{tr. note: cfr. Schouten III, (1.3)}
\end{displaymath}

\ni
We call this covariant derivative ($A_{\tau ; \alpha}$)
of $A_{\tau}$).\\
Therefore, we obtain the law of covariant differentiation

\begin{equation}
A_{\sigma;\tau} =A_{\sigma,\tau} -A_{\alpha} \Delta_{\sigma \ \tau}^{\ \alpha}
\label{e15}
\end{equation}

Analogously, out of (3) follows the formula

\begin{equation}
A^{\sigma}_{\ ;\tau} =A^{\sigma}_{\ ,\tau} +A^{\alpha} \Delta_{\alpha \ \tau}^{\ \sigma}.
\label{e16}
\end{equation}
The result is the law of covariant differentiation  for arbitrary
tensors. We illustrate this giving an example:

\begin{equation}
A_{a \ \tau;\rho}^{\ \sigma}=A_{a \ \tau,\rho}^{\ \sigma}+
A_{a \ \tau}^{\ \alpha} \Delta_{\alpha \ \rho}^{\ \sigma}-
A_{a \ \alpha}^{\ \sigma} \Delta_{\tau \ \rho}^{\ \alpha}.
\label{e17}
\end{equation}

By means  of the fundamental tensor $h_{s}^{\ \alpha}$ we are
allowed to transform  local (Latin) indices in coordinate (Greek)
indices, so we are free to favor the local or coordinate indices
when formulating some tensor relations. The first approach is
preferred by the Italian  colleagues
(Levi-Civita, Palatini), while I have preferably used coordinate indices.\\
\bf Divergence.
\rm By contraction of the covariant differential quotient
one obtains the divergence as in  the absolute differential
calculus based on metrics only. E.g., one gets  the  tensor

\begin{displaymath}
A_{\alpha \tau}=A_{\alpha\ \tau;\sigma}^{\ \sigma}.
\end{displaymath}
out of (21) by contraction of the indices $\sigma$ and $\rho$.

In earlier papers I even introduced other divergence operators,
but I do not accredit special significance to those
any more.\\
{\bf  Covariant differential quotients of the fundamental tensor.\/}\\
\rm One can  easily find out of the formulas derived above, that the covariant
derivatives and divergences of the fundamental tensor vanish. E.g.
we have

\begin{eqnarray}
h_{s \ ;\tau}^{\ \nu} \equiv h_{s \ ,\tau}^{\ \nu} +
h_s^{\ \alpha} \Delta_{\alpha \ \tau}^{\ \nu} \equiv
\delta_{s t} (h_{t \ \tau}^{\ \nu}+
 h_t^{\ \alpha}  \Delta_{\alpha \ ,\tau}^{\ \nu}) \\
\nonumber
 \equiv h_s^{\ \alpha} (h_{t \alpha}\  h_{t \ ,\tau}^{\ \nu}+
\Delta_{\alpha \ \tau}^{\ \nu}) \equiv
h_s^{\ \alpha} (-\Delta_{\alpha \ \tau}^{\ \nu}+
\Delta_{\alpha \ \tau}^{\ \nu}) \equiv 0.
\label{e18}
\end{eqnarray}

\ni
Analogously we can prove

\begin{displaymath}
h_{s  \ \tau}^{\ \nu} \equiv g^{\mu \nu}_{\ \ ;\tau}
\equiv g_{\mu \nu ;\tau} \equiv 0. \hspace{3.0cm} (18a)
\label{e18a}
\nonumber
\end{displaymath}

\ni
Likewise, the divergences $h_{s \ ; \nu}^{\ \nu}$
and $g^{\mu \nu}_{\ \ ; \nu}$ obviously vanish.\\
{\bf  Differentiation of tensor products.\/} \rm As it is apparent in the well-known
differential calculus the covariant  differential quotient
of a tensor product can be expressed by the differential quotient
of the factors. If $S_.^{\ .}$ and $T_.^{\ .}$ are tensors of arbitrary
index character,

\begin{equation}
(S_.^{\ .} \ T_.^{\ .})_{;\alpha}=S_{. \ ;\alpha}^{\ .} \ T_.^{\ .} + T_{. \ ;\alpha}^{\ .} \  S_.^{\ .} .
\label{e19}
\end{equation}

\ni
follows. Out of this and out of the vanishing covariant differential
quotient of the fundamental tensor it follows, that the latter
may be interchanged with the differentiation symbol(;).\\
{\bf  ''Curvature''\/}. \rm Out of the hypothesis of ''distant parallelism''
and out of  equation (9) we obtain the integrability of the
displacement law (10) and (11). Out of this follows

\begin{equation}
0 \equiv -\Delta_{\kappa \ \lambda; \mu}^{\ \tau}
  \equiv -\Delta_{\kappa \ \lambda, \mu}^{\ \tau}+
 \Delta_{\kappa \ \mu, \lambda}^{\ \tau}+
 \Delta_{\sigma \ \lambda}^{\ \tau} \Delta_{\kappa \ \mu}^{\ \sigma}
 -\Delta_{\sigma \ \mu}^{\ \tau} \Delta_{\kappa \ \lambda}^{\ \sigma}.
\label{e20}
\end{equation}

In order to be expressed by the entities  $h$ ,the  $\Delta$'s must
comply to these conditions (cfr.(12)). Looking at
 (20), it is clear that the characteristic laws of the manifold in
consideration here must be very different from the earlier theory.
Though according to the new theory all tensors of the earlier
theory exist, in particular the Riemannian curvature tensor calculated
from  the Christoffel symbols. But according to the new theory there are
simpler and more elementary  tensorial objects, that can be used
for formulating the field laws.\\
{\bf  The tensor $\Lambda$\/} \footnote{Cartans torsion tensor is
nowadays usually denoted as $T$ or $S$ (Schouten)}. \rm If we differentiate a scalar $\psi$
twice covariantly, we obtain according to (15) the tensor

\begin{displaymath}
\phi_{,\sigma, \tau}-\phi_{,\alpha}\ \Delta_{\sigma \ \tau}^{\ \alpha}.
\end{displaymath}

\ni
From this follows at once the  tensorial character of

\begin{displaymath}
\frac{\partial \phi}{\partial x_{\alpha}}(\Delta_{\sigma \ \tau}^{\ \alpha}-\Delta_{\tau \ \sigma}^{\ \alpha}).
\end{displaymath}

\ni
Interchanging $\sigma$ and $\tau$ a new tensor emerges and
the subtraction yields the tensor

\begin{equation}
\Lambda_{\sigma \ \tau}^{\ \alpha}=\Delta_{\sigma \ \tau}^{\ \alpha}-\Delta_{\tau \ \sigma}^{\ \alpha}.
\footnote{tr. note: cfr. Schouten III, (2.13)}
\label{e21}
\end{equation}

According to this theory there is a  tensor  containing the
components $h _{\sigma \alpha}$ of the fundamental tensor and its
first differential quotients only. We already proved that a
 vanishing fundamental tensor
causes the validity of Euclidean geometry (cfr. (13)). Therefore,
a natural law for such a continuum will consist of conditions for
this  tensor.\\
By contraction of the tensor $\Lambda$ we obtain

\begin{equation}
\phi_{\sigma}=\Lambda_{\sigma \ \alpha}^{\ \alpha}.
\footnote{tr. note: cfr. Schouten III, (2.15)}
\label{e22}
\end{equation}

A vector which, as I suspected earlier,  could take the part
of the electromagnetic  potential  in the present  theory, but
ultimately I do not uphold this view.\\
{\bf  Changing rule of differentiation.\/} \rm If a  tensor $T_.^{\ .}$
is  differentiated twice covariantly, the important rule holds

\begin{equation}
T_{. \ ;\sigma;\tau}^{\ .} - T_{. \ \tau;\sigma}^{\ .} \equiv -T_{. \ ;\alpha}^{\ .}
\Lambda_{\sigma \ \tau}^{\ \alpha}.
\label{e23}
\end{equation}

{\bf  Proof.\/} \rm If $T$ is a scalar (tensor without Greek
index), we obtain the proof without effort using (15). In this
special case
we will find the proof of the general theorem.\\
The first remark we would like to make is, that according to the theory discussed here
parallel vector fields do exist. These vector fields have the same components
in all local systems. If  ($a^{a}$)
or ($a_{a}$) is such a vector field, it satisfies the condition

\begin{displaymath}
a^{a}_{\ ;\sigma}=0 \qquad or \qquad a_{a \ ;\sigma}=0
\end{displaymath}

which can be proven easily.\\
Using such parallel vector fields the changing rule easily leads back
to the rule for a scalar.
For the sake of simplicity, we perform the proof for a  tensor $T^{\lambda}$
with only one index.
If $\phi$ is a scalar, the first thing that follows out of the definitions
(16) and (21)  is

\begin{displaymath}
\phi_{;\sigma;\tau} - \phi_{;\tau;\sigma}\equiv - \phi_{;\alpha}\
\Lambda_{\sigma \ \tau}^{\ \alpha}.
\end{displaymath}

If we put the scalar  $ a_{\lambda} T^{\lambda}$ into this
equation for $\phi$, $ a_{\lambda}$ being a  parallel vector
field, $ a_{\lambda}$ may be interchanged with the differentiation
symbol at every  covariant differentiation, therefore $
a_{\lambda}$ appears as a factor in all the terms. Therefore, one
obtains

\begin{displaymath}
[T^{\lambda}_{\ ;\sigma;\tau} - T^{\lambda}_{\ ;\tau;\sigma}+
T^{\lambda}_{\ ;\alpha} \ \Lambda_{\sigma \ \tau}^{\ \alpha}] a_{\lambda}=0.
\end{displaymath}

This identity must hold  for any choice of $ a_{\lambda}$ in a certain
position, therefore the bracket vanishes, and we have finished our proof.
 The generalization for tensors with any number of Greek
indices is obvious.\\
{\bf  Identities for the tensor $\Lambda$\/.} \rm Permuting the indices
$\kappa, \lambda, \mu$ in  (20), adding the three identities, and
by appropriate summing-up of the terms with respect to
 (21) one obtains

\begin{displaymath}
0 \equiv (\Lambda_{\kappa \ \lambda, \mu}^{\ \tau} +
          \Lambda_{\lambda \ \mu,\kappa}^{\ \tau} +
          \Lambda_{\mu \ \kappa,\lambda}^{\ \tau}) +
\Delta_{\sigma \ \kappa}^{\ \tau} \Lambda_{\lambda \ \mu}^{\ \sigma}+
\Delta_{\sigma \ \lambda}^{\ \tau} \Lambda_{\mu \ \kappa}^{\ \sigma}+
\Delta_{\sigma \ \mu}^{\ \tau} \Lambda_{\kappa \ \lambda}^{\ \sigma}.
\footnote{tr. note: cfr. Schouten III, (5.2)}
\end{displaymath}

We convert this identity by introducing covariant instead
of ordinary derivatives of the tensors $ a_{\lambda}$
(see (17)); so we acquire the identity

\begin{equation}
0 \equiv (\Lambda_{\kappa \ \lambda; \mu}^{\ \tau} +
          \Lambda_{\lambda \ \mu;\kappa}^{\ \tau} +
          \Lambda_{\mu \ \kappa;\lambda}^{\ \tau}) +
(\Lambda_{\kappa \ \alpha}^{\ \tau} \Lambda_{\lambda \ \mu}^{\ \alpha}+
\Lambda_{\lambda \ \alpha }^{\ \tau} \Lambda_{\mu \ \kappa}^{\ \alpha}+
\Lambda_{\mu \ \alpha }^{\ \tau} \Lambda_{\kappa \ \lambda}^{\ \alpha}).
\label{e24}
\end{equation}

In order to express the $ a_{\lambda}$'s
 by  the $h$ in the above manner, this condition must be satisfied.\\
Contraction of the above equation by the  indices $\tau$ and $\mu$
yields the identity

\begin{displaymath}
0 \equiv \Lambda_{\kappa \ \lambda; \alpha}^{\ \alpha} +
\phi_{\lambda;\kappa} - \phi_{\kappa ; \lambda}- \phi_{\alpha}
\Lambda_{\kappa \ \lambda}^{\ \alpha}.
\end{displaymath}

\ni
or

\begin{equation}
\Lambda_{\kappa \ \lambda; \alpha}^{\ \alpha} +
\phi_{\lambda,\kappa} - \phi_{\kappa ,\lambda}
\label{e25}
\footnote{tr. note: cfr. Schouten III, (5.6)}
\end{equation}

\ni
where  $\phi_{\lambda}$ stands for
$\Lambda_{\lambda \ \alpha}^{\ \alpha}$  (22).

\section{$\S~4.$ The field equations}

The  most simple field equations we desired to find will be conditions for the
 tensor $\Lambda_{\lambda \ \nu}^{\ \alpha}$ . The number of
 $h$-components is $n^2$, of which $n$
remain indeterminate due to general covariance; therefore the number of independent
field equations must be $n^2 - n$.
On the other hand, the higher number of possibilities a theory cuts down on
(without contradicting experience), the more satisfactory it is.
Therefore, the number  $Z$ of  field equations should be as large
as possible. If $\bar Z$ denotes the number of identities
in-between the field equations, $Z-\bar Z$ must be equal to $n^2 -n$.\\
According to the change rule of differentiation

\begin{equation}
\Lambda_{\mu \ \nu; \nu; \alpha}^{\ \alpha}-
\Lambda_{\mu \ \nu; \alpha; \nu}^{\ \alpha}-
\Lambda_{\mu \ \tau; \alpha}^{\ \sigma}\
\Lambda_{\sigma \ \tau}^{\ \alpha} \equiv 0.
\footnote{tr. note: cfr. Schouten III, (4.9)}
\label{e26}
\end{equation}

\ni
holds. An underlined index indicates  ''pulling up''
and ''pulling down'' of an  index, respectively, e.g.

\begin{displaymath}
\Lambda_{\underline{\mu} \  \underline{\nu}}^{\ \alpha} \equiv
\Lambda_{\beta \ \gamma}^{\ \alpha} \ g^{\mu \beta} \ g^{\nu \gamma},
\end{displaymath}

\begin{displaymath}
\Lambda_{\mu \ \nu}^{\ \underline{\alpha}} \equiv
\Lambda_{\mu \ \nu}^{\ \beta}\  g_{\alpha \beta}.
\end{displaymath}

\ni
We write this Identity (26)  in the form

\begin{displaymath}
G^{\mu \alpha}_{\ \ ; \alpha}-F^{\mu \nu}_{\ \ ; \nu}+
\Lambda_{\underline{\mu} \ \underline{\tau}}^{\ \sigma} F_{\sigma \tau} \equiv 0,
\hspace{3.0cm} (26a)
\label{e26a}
\end{displaymath}

\ni
with the following settings

\begin{equation}
G^{\mu \alpha} \equiv \Lambda_{\underline{\mu} \ \underline{\nu} ; \nu}^{\ \alpha}
-\Lambda_{\underline{\mu} \ \underline{\tau}}^{\ \sigma}
\Lambda_{\sigma \ \tau}^{\ \alpha},
\label{e27}
\end{equation}

\begin{equation}
F^{\mu \nu} \equiv \Lambda_{\underline{\mu} \ \underline{\nu} ;\  \alpha}^{\ \alpha}.
\label{e28}
\end{equation}

\ni
Now we make an ansatz for the field equations:

\begin{equation}
G^{\mu \alpha}=0,
\label{e29}
\end{equation}

\begin{equation}
F^{\mu \alpha}=0.
\label{e30}
\end{equation}

These equations seem to contain an forbidden overdetermination,
because their number is $n^2 + \frac{n (n-1)}{2}$, while at first hand
it is only known to satisfy the identities (26a).\\
Linking (25) with (30) it follows,
that the  $\phi_k$ can be derived from a potential.
Therefore, we set

\begin{equation}
F_{\kappa}=\phi_{\kappa}-\frac{\partial \log \psi}{\partial x^{\kappa}} =0.
\label{e31}
\end{equation}

(31) is completely equivalent with  (30). The equations
(29), (31) combined are  $n^2 +n$ equations for
 $n^2+1$ functions $h_{s \nu}$ and $\psi$. Besides  (26a)
there is, however, another system of identities between these equations
we will derive now.\\
If $\underline G^{\mu \alpha}$ denotes the antisymmetric
part of  $G^{\mu \alpha}$, one can figure out directly from
 (27)

\begin{equation}
2 \  {\underline{G}}^{\mu \alpha}=
S_{\mu \ \alpha;\nu}^{\ \nu}
+\frac{1}{2} S_{\underline{\sigma} \ \underline{\tau}}^{\ \mu}
\Lambda_{\sigma \ \tau}^{\ \alpha}
-\frac{1}{2} S_{\underline{\sigma} \ \underline{\tau}}^{\ \alpha}
\Lambda_{\sigma \ \tau}^{\ \mu}+
F^{\mu \alpha},
\label{e32}
\end{equation}

\ni
For the sake of abbreviation  we introduce the totally  skew-symmetric tensor

\begin{equation}
S_{\underline{\mu} \ \underline{\nu}}^{\ \alpha}=
\Lambda_{\underline{\mu} \ \underline{\nu}}^{\ \alpha}+
\Lambda_{\underline{\alpha} \ \underline{\mu}}^{\ \nu}+
\Lambda_{\underline{\nu} \ \underline{\alpha}}^{\ \mu}.
\label{e33}
\end{equation}

\ni
Figuring out the first term of  (32)
yields

\begin{equation}
2 \ \underline{G}^{\mu \alpha}=
S_{\underline{\mu} \ \underline{\alpha};\nu}^{\ \nu}
- S_{\underline{\mu} \ \underline{\alpha}}^{\ \sigma}
\Lambda_{\sigma \ \nu}^{\ \nu}+
F^{\mu \alpha},
\label{e34}
\end{equation}

\ni
But now, with respect to the definition of $F_k$ (31)

\begin{displaymath}
\Delta_{\sigma \  \nu}^{\ \nu}-\Delta_{\nu \  \sigma }^{\ \nu}
\equiv \Lambda_{\sigma \  \nu}^{\ \nu} \equiv \phi_{\sigma}
\equiv F_{\sigma}+\frac{\partial \log \psi}{\partial x^{\sigma}}
\end{displaymath}

\ni
or

\begin{equation}
\Delta_{\sigma \  \nu}^{\ \nu}= \frac{\partial \log \psi \ h}{\partial x^{\sigma}}
+F_{\sigma}
\label{e35}
\end{equation}

\ni
holds. Therefore, (34) takes the form

\begin{displaymath}
h \psi (2 \ \underline G^{\mu \alpha} - F^{\mu \alpha} +
S_{\underline \mu \  \underline {\alpha}}^{\ \sigma} F_{\sigma})
\equiv \frac{\partial}{\partial x^{\sigma}}
(h \  \psi S_{\underline \mu \ \underline {\alpha}}^{\ \sigma})  \qquad \qquad \qquad  \qquad \qquad \qquad (34b)
\label{e34a}
\end{displaymath}

\ni
Due to the antisymmetry the desired system of identical
equations follows

\begin{equation}
\frac{\partial}{\partial x^{\alpha}} [ h \ \psi (2 \ \underline G^{\mu \alpha} - F^{\mu \alpha} +
S_{\underline \mu \  \underline \alpha}^{\ \sigma} F_{\sigma}) ]
\equiv 0
\label{e36}
\end{equation}

These are at first $n$ identities, but only  $n-1$ of them are
linear independent from each other. Because of the antisymmetry
$[\ ],_{\alpha, \mu} \equiv 0$ holds independently no matter what one inserts
in  $G^{\mu \alpha}$ and $F_{\mu}$.\\
In the identities (4) and (36) you have to think of $F^{\mu
\alpha}$ being expressed by $F_{\mu}$  according to the following
relation which was derived from (31)

\begin{displaymath}
F_{\mu \alpha} \equiv F_{\mu, \alpha}- F_{ \alpha,\mu}.  \hspace{3.0cm} (31a)
\label{e31a}
\end{displaymath}

Now we are able to prove the compatibility  of the field equations
(29), (30) or   (29), (31), respectively.\\
First of all we have to show that the number of field equations
minus the number of (independent) identities is smaller by $n$
than the number of field variables. We have

\begin{quote}
number of field equations (29) (31) :\qquad $n^2+n$\\
number of (independent) identities:\qquad  $n+n-1$\\
number of field variables:\qquad  $n^2+1$, \\
$(n^2+n)-(n+n-1)=(n^2+1)-n$
\end{quote}

As we see the  number of identities just fits. We do not
stop here, but prove the following\\
{\bf  Proposition\/}. {\em If in a cross section $x^n =const.$ all
 differential equations are satisfied and, in addition, $(n^2+1)-n$
of them are  properly chosen everywhere, then all $n^2+n$ equations
are fulfilled anywhere.\/}\\
\rm
{\bf  Proof\/}.\rm  If all equations are fulfilled in the cross section  $x^n=a$
and if these equations, that correspond to setting to zero the
below, are fulfilled everywhere, we obtain:

\begin{displaymath}
\begin{array}{cccc}
F_{1}  &\ldots & F_{n-1} F_n \\
G^{1 \ 1} & \ldots &  G^{1 \ n-1} \\
\dots \\
G^{n-1 \  1} & \ldots & G^{n-1 \  n-1}. \\
\end{array}
\end{displaymath}

 Then from (4) follows, that the
$F^{\mu \alpha}$ vanish everywhere. Now one deduces from (36),
that in an neighboring cross section $x^n=a+d a$ the
 skew-symmetric  $\underline G^{\mu \alpha}$ for $\alpha=n$
must vanish as well~\footnote{The
$\frac{\partial G^{\mu n}}{\partial x^{n}}$ vanish for
$x^n=a$.}. Out of (26a) it then follows, that in addition
the symmetric  $\underline G^{\mu \alpha}$ for  $\alpha=n$
at the  adjacent cross section  $x^n=a+d a$ must vanish.
Repeating this kind of deduction proves the proposition.

\section{$\S~5.$ First approximation}

We are now going to deal with a field that shows very little
difference from an  Euclidean one with  ordinary  parallelism.
Then we may set

\begin{equation}
h_{s \nu} =\delta_{s \nu}+ \bar h_{s \nu},
\label{e37}
\end{equation}

\ni
where $\bar h_{s \nu}$ is infinitely small at first order, higher
order terms are neglected.
Then, according to (5) and (6), we have to set

\begin{equation}
h_{s}^{\ \nu} =\delta_{s \nu}- \bar h_{\nu s}.
\label{e38}
\end{equation}

\ni
In first approximation, the  field equations (29), (31)
read

\begin{equation}
\bar h_{a \mu,\  \nu,\  \nu} - \bar h_{a \nu, \ \nu,\  \mu} =0,
\label{e39}
\end{equation}

\begin{equation}
\bar h_{a \mu,\  a,\  \nu} - \bar h_{a \nu, \ a, \ \mu} =0.
\label{e40}
\end{equation}

\ni
we substitute equation (31) by

\begin{displaymath}
\bar h_{a \nu, a } = \chi_{\nu}.   \qquad \qquad  \qquad (40a)
\label{e40a}
\end{displaymath}

We claim now that there is an infinitesimal coordinate transformation
$x^{\nu'} =x^{\nu}- \xi^{\nu}$, which causes all the variables
$\bar h_{\alpha \nu , \nu}$ und $\bar h_{\alpha \nu , \alpha}$
to vanish.\\
{\bf  Proof.\/} \rm First we prove that

\begin{equation}
\bar h'_{\mu \nu} = \bar h_{\mu \nu }\  \xi^{\mu}_{\ , \nu},
\label{e41}
\end{equation}

\ni
Therefore,

\begin{displaymath}
\bar h'_{a \nu, \nu} = h_{a \nu, \ \nu }+ \xi^{\alpha}_{\ , \ \nu, \ \nu},
\end{displaymath}

\begin{displaymath}
\bar h'_{a \nu a} = \bar h_{a \nu  a} + \xi^{\alpha}_{\ , \ a \nu}.
\end{displaymath}

\ni
The right sides vanish because of  (40a), if
the following equations are fulfilled

\begin{eqnarray}
\xi^{\alpha}_{\ ,\ \nu,\  \nu}= -\bar h_{a \nu,\  \nu},\\
\nonumber
\xi^{\alpha}_{\ ,\ a}= -\chi.
\label{e42}
\end{eqnarray}

\ni
But these $n+1$ equations for  $n$ variables $\xi_{\alpha}$ are
compatible, because of (40a)

\begin{displaymath}
(-\bar h_{a \nu, \nu})_{,\ a} -(-\chi)_{,\ \nu ,\ \nu }=0.
\end{displaymath}

\ni
Choosing new coordinates, the field equations read

\begin{eqnarray}
\bar h_{a \mu,\  \nu \nu} =0,\\
\nonumber
\bar h_{a \mu,\  a} =0,\\
\nonumber
\bar h_{a \mu,\  \mu} =0,\\
\nonumber
\end{eqnarray}

\ni
If we now separate $\bar h_{\alpha \nu , \nu}$ according to the equations

\begin{eqnarray}
\nonumber
\bar h_{a \mu} +\bar h_{\mu a} =\bar g_{a \mu} ,\\
\nonumber
\bar h_{a \mu} -\bar h_{\mu a} = a_{a \mu} ,\\
\nonumber
\end{eqnarray}

\ni
where $\delta_{\alpha \mu} +\bar g_{\alpha \mu} (=g_{\mu \nu})$
determines  metrics in first approximation , thus the field
equations take the simple form

\begin{equation}
\bar g_{a \mu,\  \sigma,\  \sigma}=0,
\label{e44}
\end{equation}

\begin{equation}
\bar g_{a \mu,\ \mu}=0,
\label{e45}
\end{equation}

\begin{equation}
 a_{a \mu,\  \sigma,\  \sigma}=0,
\label{e46}
\end{equation}

\begin{equation}
a_{a \mu,\ \mu}=0.
\label{e47}
\end{equation}

One is led to suppose, that $\bar g_{\alpha \nu}$ and
 $a_{\alpha \mu}$ represent the gravitational and the
 electromagnetic  field in first approximation respectively.
(44), (45) correspond to Poisson's equation,
(46), (47) to Maxwell's
equations of the empty space. It is interesting that the field laws
of  gravitation seem to be separated from those of the electromagnetic
field, a fact which is in agreement  with the observed independence of the two fields.
But in a strict sense none of them exists separately.\\
Regarding the covariance of the equations (44) to (47)
we note the following. For the $h_{s \mu}$'s generally the transformation
law

\begin{displaymath}
 h^{'}_{\ s \mu}= \alpha_{s t}
\frac{\partial x^{\sigma}}{\partial x^{\mu '}}\  h_{t \sigma}
\end{displaymath}

\ni
holds. If the coordinate transformation is chosen linear and
orthogonal as well as conform with respect to the twist of
the local systems, that is

\begin{equation}
 x^{\mu '}= \alpha_{\mu \sigma} x^{\sigma},
\label{e48}
\end{equation}

\ni
we acquire the transformation law

\begin{equation}
 h'_{s \mu}= \alpha_{s t}\  \alpha_{\mu \sigma}\  h_{ t \sigma},
\label{e49}
\end{equation}

\ni
which is exactly the same as for tensors in special relativity.
Because of (48) the same transformation law holds  for the
 $\delta_{s \mu}$, so it  also holds for the  $\bar h_{\alpha \mu},
\bar g_{\alpha \mu}$,
and $a_{\alpha \mu}$. With respect to such transformations the
 equations (44) to (47) are covariant.

\section{$\S~6.$ Outlook}

The big appeal of the theory exposed here, lies in its unifying
structure
 and the high-level (but allowed) overdetermination of the
field variables. I was able to show that the field equations yield
equations, in first-order approximation, that correspond to the
Newton-Poisson theory of gravitation and to Maxwell's theory of the
electromagnetic field. Nevertheless I'm still far away from
 claiming the physical validity of the equations I derived. The
reason for that is, that I did not succeed in deriving equations
of motion for particles yet.

\end{document}